\pdfoutput=1 
\documentclass[twocolumn,aps,prl,nofootinbib]{revtex4-1}
\usepackage{graphicx}
\usepackage{amsmath}
\usepackage{amssymb}
\def\ben{\begin{equation}}
\def\een{\end{equation}}
\def\sss{\scriptscriptstyle\rm}
\def\s{_{\sss S}}
\def\xc{_{\sss XC}}
\def\nc{\left<n_{\sss C}\right>}
\def\tchic{\tilde{\chi}_{\rm c}}

\begin{document}

\title{Accuracy of density functionals for molecular electronics: the Anderson junction}
\author{Zhen-Fei Liu, Justin P. Bergfield, and Kieron Burke}
\affiliation{Departments of Chemistry and of Physics,
University of California, Irvine, California 92697, USA}
\author{Charles A. Stafford}
\affiliation{Department of Physics, University of Arizona, 1118 East Fourth Street, Tucson, Arizona 85721, USA}
\date{\today}

\begin{abstract}
The exact ground-state exchange-correlation functional
of Kohn-Sham density functional theory yields the
exact transmission through
an Anderson junction at zero bias and temperature. 
The exact impurity charge susceptibility is used to construct the exact exchange-correlation potential. We analyze the successes and limitations of various types of approximations, including smooth and discontinuous functionals of the occupation, as well as symmetry-broken approaches.
\end{abstract}

\maketitle

Since the pioneering experiments of Reed and Tour on dithiolated benzene \cite{RZMB97},
there has been tremendous progress in the ability to both create and
characterize \cite{CFR05} organic molecular junctions.  But accurate simulation of
these devices remains a challenge, both theoretically and computationally \cite{NR03}.
The essential physics has been well understood since the
ground-breaking work of Landauer and B\"{u}ttiker \cite{L57,B86}
in the context of mesoscopic devices,
including both Coulomb blockade and Kondo effects \cite{MWL91,MWL93}.
Calculations with simple model Hamiltonians demonstrate such
effects at a qualitative level \cite{L95}.  
On the other hand, organic molecules connected to metal leads \cite{EWK04}
require hundreds of atoms and thousands of basis functions for a
sufficiently accurate calculation of their total energy, geometry,
and single-particle states.  Such conditions are routine
for modern density functional theory (DFT) calculations \cite{primer},
but the ability of present functional approximations
 to predict accurate currents remains an open question \cite{KCBC08}.

The standard DFT method for calculating current through such a
device is to perform a ground-state
Kohn-Sham (KS) DFT calculation \cite{KS65} on a system upon
which a difference between the chemical potentials
of the left and right leads has been imposed (the applied bias), and
calculate the transmission through the KS potential using the Landauer-B\"{u}ttiker formula.
But there is nothing in the basic theorems of DFT that directly implies
that such a calculation would yield the correct current, even
if the exact ground-state functional were used.

The limit of weak bias is more easily analyzed than the general case,
because the Kubo linear response formalism applies \cite{FL81,BS89}. 
In that case one finds that, in principle, there are exchange-correlation (XC)
corrections to the current in the standard approach \cite{KBE06}, 
but little is known about their magnitude \cite{SZVD05,JBG07}.  Even without these corrections,
one can ask if the standard approximations used in most ground-state
DFT calculations (i.e., generalized gradient approximations \cite{PBE96}
and hybrids of these with Hartree-Fock exchange \cite{B93,PEB96})
are sufficiently accurate for transport purposes. The answer appears definitively no!
Because of self-interaction errors, such approximations are well-known \cite{PZ81} 
to produce potentials with incorrectly positioned KS eigenvalues, both
occupied and unoccupied. These errors become severe when the molecule is
only weakly coupled to the leads \cite{TFSB05}.  Calculated transmission can be
too large by several orders of magnitude due to
this incorrect positioning of the levels.  Recent calculations 
\cite{SE08} using beyond-DFT techniques to correctly position the levels
show greatly improved agreement with experiment.

But this progress returns us to the earlier concern:  Even with an exact ground-state
XC functional, are there XC corrections to the Landauer-B\"{u}ttiker result?
The answer appears to be yes in general \cite{KBE06},
but in a previous work \cite{BLBS11} we argued that, under a broad
range of conditions applicable to typical experiments, such
corrections can vanish. This result was shown by exact calculations on
an impurity model (Anderson model) employing the exact XC functional.
In the present work, we analyze different approximate treatments, applied to the
Anderson junction, and calculate their errors.
The implications for DFT calculations of transport in general are discussed. 

\begin{figure}
\begin{center}
\includegraphics[width=3.5in]{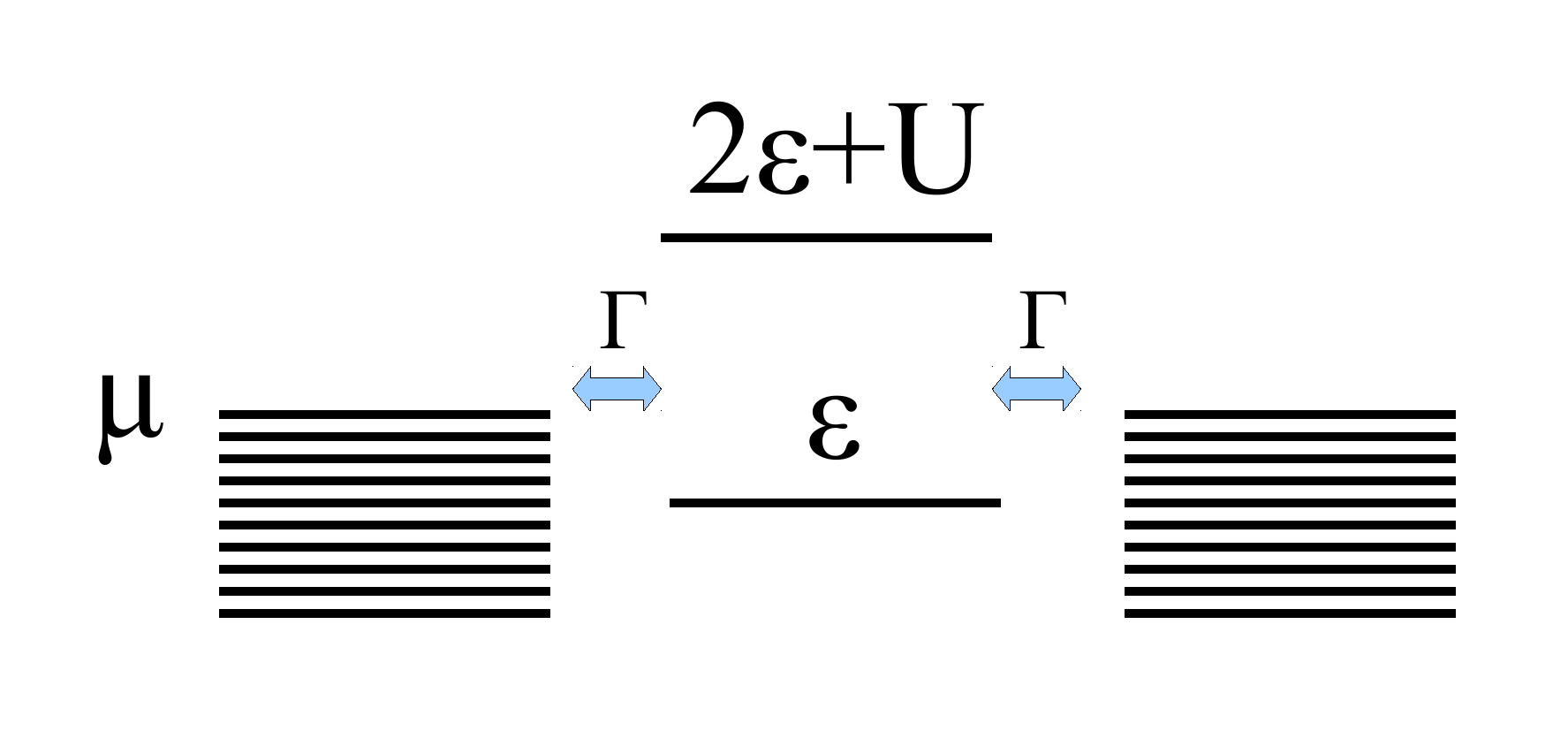}
\caption{A cartoon for Anderson model. The model consists of two featureless leads and a central region with on-site interaction $U$. $\Gamma$ is the tunneling width. Two many-body levels of the central region are shown.}
\label{f:cartoon}
\end{center}
\end{figure}

The Anderson model \cite{A61} is a single interacting site (C) 
connected to two non-interacting electrodes (L,R). The Hamiltonian of the
system is ${\cal H}={\cal H}_{\rm C}+{\cal H}_{\rm T}+{\cal H}_{\rm L,R}$.
Each lead is represented by a non-interacting Fermi gas: 
${\cal H}_{\rm L,R}=\sum_{k\sigma \in L,R}\varepsilon_{k\sigma} {\hat n}_\sigma$, 
with chemical potential $\mu$
and the central interacting site is: 
${\cal H}_{\rm C}=\varepsilon\left({\hat n}_\uparrow+{\hat n}_\downarrow\right)+U{\hat n}_\uparrow {\hat n}_\downarrow$,
where ${\hat n}_\sigma=d^\dagger_\sigma d_\sigma$ is the number
operator for spin $\sigma$ and $U$ is the charging energy
representing on-site interaction. ${\cal H}_{\rm T}$ is the
tunneling between leads and the central site.
The tunneling width $\Gamma$ is a constant in the broad-band limit.
A schematic is shown in Fig. \ref{f:cartoon}.  
Real molecules can be mapped onto the Anderson model \cite{BS09,BSSR11}.

In a previous work \cite{BLBS11}, we calculated the exact relation
between occupancy on the central site and on-site energy $\varepsilon$ for an Anderson junction, using the Bethe ansatz (BA) \cite{WT83}.
We showed that exact KS DFT yields the exact transport at zero temperature
and in the linear response regime, although the KS spectral function differs
from the exact one away from the Fermi energy. This is because the Anderson junction
has only one site and transmission is a function of occupation number
due to the Friedel-Langreth sum rule \cite{F58,L66}. 
Thus, for this simple model, all failures of approximate XC calculations
of transmission can be attributed to failures to reproduce
the exact occupation number, i.e., there are {\em no} XC corrections to 
the standard practice of applying KS DFT to the ground-state and finding
transmission through the single-particle potential.
On the other hand, the standard approximations in use in DFT calculations
of transport have a variety of shortcomings.  The most prominent one, as we
shall see, is the 
lack of a discontinuity in the XC potential with particle number \cite{PPLB82}.

Before studying approximations,
we refine our previous numerical fit of
BA results, using analytic results from many-body
theory.  We re-introduce \cite{A61}
reduced variables $y=\Gamma/U$, which measures the ratio of lead-coupling
to the onsite Coulomb repulsion, while $x=(\mu-\epsilon)/U$
is the difference between the leads'
chemical potential and the onsite level energy, in units of $U$.
For $x< 0$, the central site
is above the chemical potential, at $x=0$ they match.

The occupation in the KS system is given by
self-consistent solution of the KS equation for occupation:
\ben
\nc = \frac{1}{2}+\frac{1}{\pi}\arctan \left( \frac{\mu-\varepsilon\s(\nc)}{\Gamma}\right).
\label{nmu}
\een
where the KS level is written as
\ben
\varepsilon\s(\nc) = \varepsilon + \frac{U}{2}\nc + \varepsilon\xc(\nc),
\label{kses}
\een
with the second term being the Hartree contribution and the third being the XC contribution
(in fact, only correlation, as exchange is zero for this model), which is a function of
the occupation.
Considered in reverse, this is a {\em definition} of the exact $\varepsilon\xc$, if the occupation
is known, as it is from the BA solution.  
The KS transmission is then
\ben
T(E)_{E=\mu}=\sin^2\left(\frac{\pi}{2}\nc\right),
\label{tnospin}
\een
and matches the {\em true} transmission in the many-body system, by virtue of the sum-rule. The exact ground-state functional yields the exact transmission, including the Kondo plateau at zero temperature and weak bias \cite{PG04,BLBS11}.

As shown in Ref. \cite{BLBS11}, the XC potential can be very accurately parametrized with the form:
\ben
\frac{\varepsilon\xc}{U}=
\frac{\alpha}{2}\left[1-\nc-\frac{2}{\pi}\tan^{-1}\left(\frac{1-\nc}{\sigma}\right)\right]
\label{justinfit}
\een
The $\tan^{-1}$ term jumps by $\pi$ as $\nc$ passes through 1, leading to discontinuous
behavior with occupation.  Thus $\sigma$ determines the width of this region, while
$\alpha$ determines its strength.   Both $\sigma$ and $\alpha$ are functions of
$y=\Gamma/U$ and were extracted numerically by fitting to the exact solution, and were
roughly fit by simple Pad\'e approximations there.

\begin{figure}
\begin{center}
\includegraphics[width=3.5in,trim=0.5in 0.5in 0.5in 0.5in,clip=true]{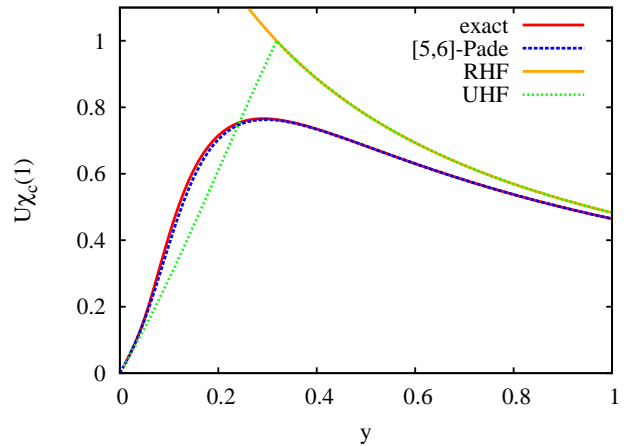}
\caption{Dimensionless susceptibility $\tchic=U\chi_{\rm c}(1)$ as a function of $y=U/\Gamma$ for the
Anderson junction [exact, [5,6]-Pad\'{e} fit (see text), RHF and UHF].}
\label{fig:exactcond}
\end{center}
\end{figure}

\begin{figure}
\begin{center}
\includegraphics[width=3.5in,trim=0.5in 0 0.4in 0,clip=true]{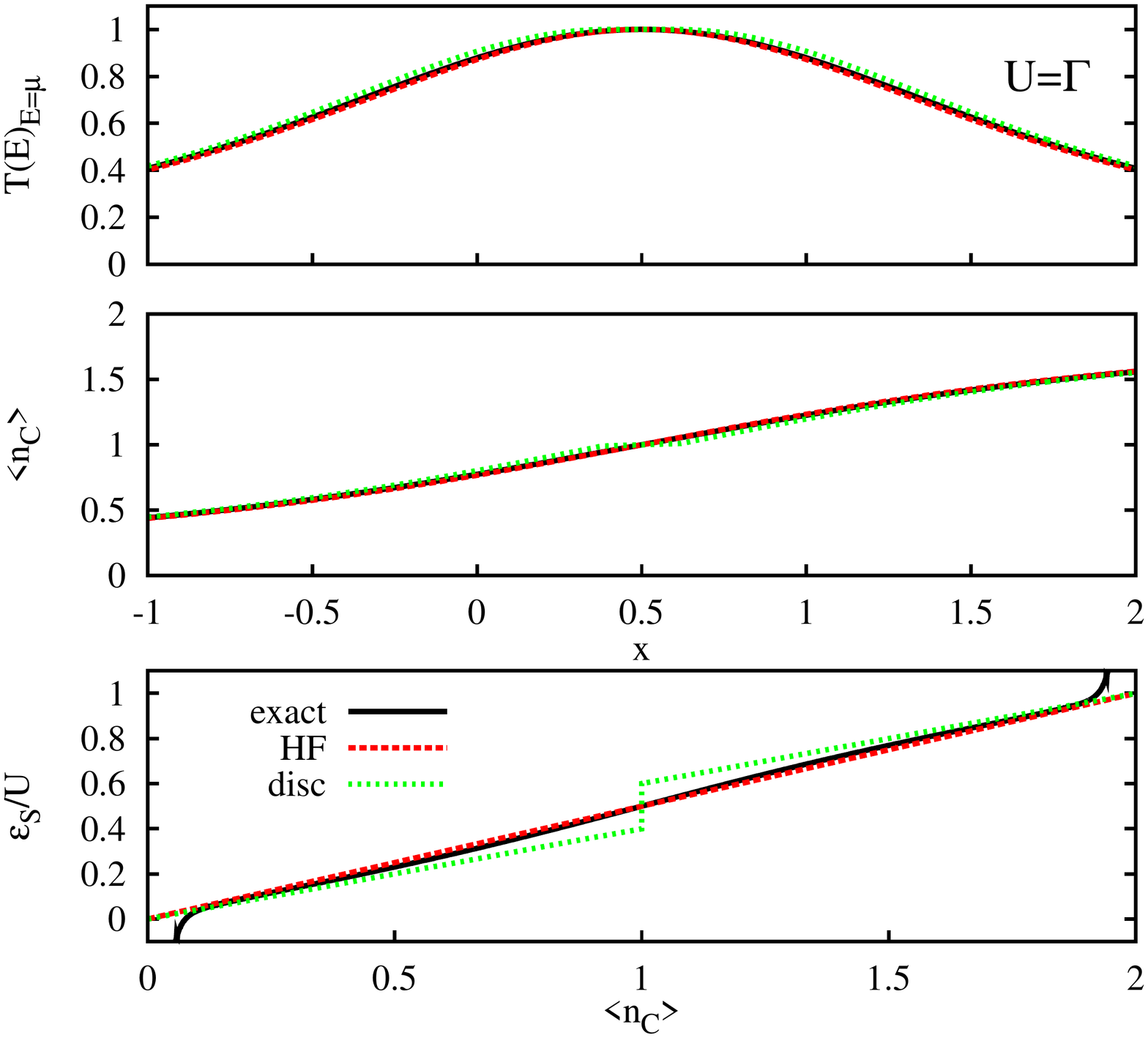}
\caption{Upper panel: transmission as a function of $x=(\mu-\epsilon)/U$; middle panel: occupation as a function of $x$; lower panel: KS potential as a function of occupation. Results are shown for Bethe ansatz or exact KS DFT (exact), Hatree-Fock (HF), and discontinuous approximation [disc, Eq. \eqref{ldau}]. $U=\Gamma$ in all cases.}
\label{f:u1g1}
\end{center}
\end{figure}

\begin{figure}
\begin{center}
\includegraphics[width=3.5in,trim=0.5in 0 0.4in 0,clip=true]{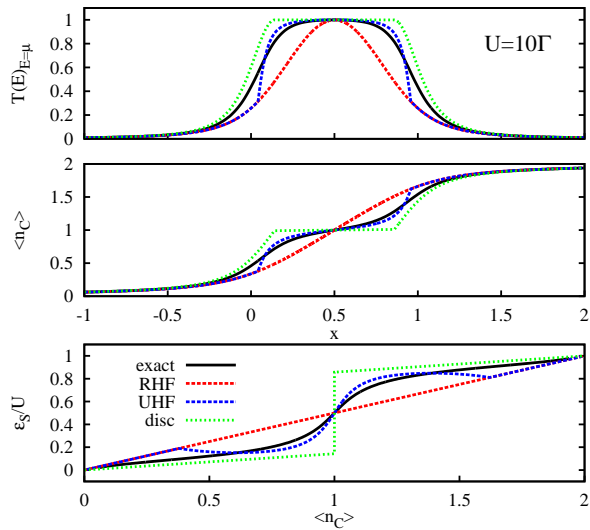}
\caption{Upper panel: transmission as a function of $x=(\mu-\epsilon)/U$; middle panel: occupation as a function of $x$; lower panel: KS potential as a function of occupation. Results are shown for Bethe ansatz or exact KS DFT (exact), restricted Hatree-Fock (RHF), unrestricted Hartree-Fock (UHF), and discontinuous approximation [disc, Eq. \eqref{ldau}]. $U=10\Gamma$ in all cases.}
\label{f:u10g1}
\end{center}
\end{figure}

However, we can greatly improve the fit of $\sigma$. A central object in the Anderson junction is the charge susceptibility, $\chi_{\rm c}(\nc) = d\nc/d\mu $. At half-filling, this is known analytically\cite{ZH83,SJG09}:
\ben
\tchic=
\frac{1}{\pi}\sqrt{\frac{2}{y}}\int_{-\infty}^{\infty}dt\, \frac{e^{-\pi y t^2/2}}
{1+\left({(2y)}^{-1}+t\right)^2},
\label{uchi}
\een
where $\tchic = U\chi_{\rm c}(1)$ is dimensionless, and is plotted in Fig. \ref{fig:exactcond}.
This curve can be readily fit to a [5,6] Pad\'e form:
\ben
\tchic^{\rm mod}(y)=\sum_{k=1}^5 a_k y^k/\sum_{k=0}^6 b_k y^k,
\label{padeform}
\een
whose 11 independent
coefficients are chosen to recover the Taylor-expansion around $y=0$ (strongly-correlated
limit)\footnote{In Refs. \cite{H97} and \cite{ES11}, this expansion was reported incorrectly, with minus sign on the second term. We believe it is corresponding to Wilson ratio $R=1$, however this is not true in strongly-correlated limit.} exactly
to 5 orders, around $y\to \infty$ to 6 orders, and are given in Table \ref{tab:padecoeff}.
The weak-correlation limit can also be extracted via the
Yosida-Yamada perturbative approach \cite{YY70,Y75,YY75}.
The quantity $\tchic$ has the physical meaning of the
slope at particle-hole symmetry point in the $\nc$ 
vs. $(\mu-\epsilon)/U$ curve (see Figs. \ref{f:u1g1} and \ref{f:u10g1}). 
It has a maximum at about $y=0.291$, and as $y$ varies from $\infty$ (weakly-correlated limit) to
$0$ (strongly-correlated limit), the slope at the symmetric point increases at first.
Beyond the maximum value, the slope decreases, and the 
Coulomb blockade pleateau gradually develops.

\begin{widetext}

\begin{table}[h]
\caption{Coefficients in the [5,6]-Pad\'{e} approximation [Eq. \eqref{padeform}].}
\begin{center}
\begin{tabular}{c|c c}
\hline\hline
$k$ & $a_k$ & $b_k$ \\
\hline
0 & $-$ & $\pi^3(\pi^6+6\pi^4-225\pi^2+675)$\\
1 & $8\pi^2$ & $-12\pi^2(\pi^6+54\pi^4-945\pi^2+3105)$ \\
2 & $-576\pi(8\pi^4-120\pi^2+405)$ & $12\pi(\pi^8-30\pi^6+555\pi^4-6525\pi^2+29700)$ \\
3 & $64(\pi^8-36\pi^6+153\pi^4+135\pi^2+8910)$ & $96(\pi^8-80\pi^6+975\pi^4-2925\pi^2+1350)$ \\
4 & $256\pi(4\pi^6-204\pi^4+1530\pi^2+945)$ & $48\pi(\pi^8-30\pi^6-225\pi^4+3375\pi^2+8100)$ \\
5 & $128\pi^2(\pi^6-42\pi^4+315\pi^2+135)$ & $576\pi^2(\pi^6-50\pi^4+375\pi^2+225)$ \\
6 & $-$ & $\pi a_5/2$ \\
\hline\hline
\end{tabular}
\end{center}
\label{tab:padecoeff}
\end{table}
\end{widetext}

By taking derivatives on both sides of Eq. \eqref{justinfit}, the two coefficients $\alpha$ and $\sigma$ 
are constrained by $\tchic$:
\ben
\sigma=\frac{2\alpha}{\pi\left( 2/ \tchic-y\pi+\alpha-1\right)}.
\label{alfsig}
\een
Retaining the simple form of Ref. \cite{BLBS11}, a [0,1] Pad\'e, 
$\alpha=1/(1+5.68y)$, we determine $\sigma$ from the Pad\'e fit to
the susceptibility and Eq. \eqref{alfsig}.
This yields highly accurate occupations, KS potentials, and transmissions, including
the Kondo plateau.  It agrees very well with the numerical fit
to the BA results of Ref. \cite{BLBS11}, and matches more closely than the 
simpler analytic fit
used there.

We now move on to the central topic of this work, which is the accuracy of approximate
functional treatments.  In such treatments, $\varepsilon\xc$ is approximated as a function of
$\nc$ in Eq. \eqref{kses}, and the resulting Eq. \eqref{nmu} is solved self-consistently for $\nc$.
The simplest such approximation is to simply set $\varepsilon\xc=0$, i.e., Hartree-Fock (HF), and
should be accurate when correlation is weak.  In Fig. \ref{f:u1g1}, we plot several quantities
for $U=\Gamma$, both exactly and in HF, showing that HF is very accurate here. 
We find \cite{A61}:
\ben
\tchic^{\rm HF}=\frac{2}{1+y\pi},
\een
which is correct to leading order in $y^{-1}$:
\ben
\tchic \to 2/(\pi y)-2/(\pi y)^2+2\gamma/(\pi y)^3+\cdots \hspace{0.2in} y \to \infty,
\een
where $\gamma=3-\pi^2/4$ exactly, but $\gamma=1$ in HF. Thus we regard $U \lesssim \Gamma$ as the weakly correlated regime.
On the other hand, in Fig. \ref{f:u10g1}, we show the same plots for $U=10 \, \Gamma$.  Now, in the exact
occupation, the slope near $x=0.5$ is much weaker, leading to a transmission plateau
(the Kondo plateau) for $ 0 \leq x \leq 1$.  The plateau effect is missed entirely by
HF, because of the too-smooth dependence (in fact, linear)
of its KS level on occupation (see bottom
panel).  Note that at temperatures equal to or above the Kondo
temperature, the Kondo effect is destroyed, and the central plateau in transmission is replaced by two
Hubbard peaks around $x=0$ and $1$.  Then the behavior of the HF curve is exactly
as qualitatively predicted in Ref. \cite{KBE06}, smearing out the two sharp features into
one peak midway between them.  This is because the KS level shifts linearly with 
occupation in HF, instead of more suddenly with occupation in the exact solution.
More generally, all smooth density functionals, such as the local density approximation \cite{KS65}
and the generalized gradient approximation \cite{PBE96}, suffer from the same qualitative
failure, and so would produce incorrect peaks centered at $x=0.5$.
All these errors arise from the approximations to the functional; the exact ground-state
functional reproduces the exact occupation by construction, and so yields the
exact transmission.

There have thus been several suggestions \cite{LSOC03} to incorporate the discontinuous behavior
with occupation into approximations in transport calculations.  At the practical level, Toher et al. \cite{TFSB05} showed in a model 
calculation how self-interaction corrections would greatly suppress zero-bias conductance
in local density approximation calculations for molecules weakly coupled to leads.  More recently, Bethe Ansatz Local Density Approximation (BALDA) \cite{LSOC03} and
variations \cite{KSKV10} have been used to impose discontinuous behavior on the levels.
For simple models, all of these can be considered as LDA$+U$-like.
The methodology of LDA$+U$ \cite{AZA91} has become increasingly popular in recent years,
especially for those focused on moderately correlated systems such as transition
metal oxides, for which LDA and GGA often have zero KS band gap.
In some fashion, a Hubbard $U$ is added to some orbitals of a DFT Hamiltonian.
Sometimes $U$ is regarded as an empirical parameter, while others have found
self-consistent prescriptions.  In any event, despite not fitting in the
strict DFT framework, it is a method borne of practical necessity for many situations \cite{KCSM06}.

To gain a qualitative understanding of the effects of such models, we define
a very simple XC potential that has a discontinuity.
To do this,
we simply take the Hartree form, symmetrize it around the half-filled point, and 
replace $U$ by a screened $\tilde U$. We find that a simple fit $\tilde U=U/(1+0.25/y)$ works well. $\tilde U$ being different from $U$ and particle-hole symmetry guarantee an explicit
derivative discontinuity of $\varepsilon\s$ with respect to occupation number.
This yields
\ben
\varepsilon\s[n]=\frac{1}{2} \tilde U n \theta(1-n)
+\left[U+\frac{1}{2}\tilde U (n-2)\right] \theta(n-1),
\label{ldau}
\een
where $\theta(x)$ is the Heaviside theta function 
and for simplicity, $n$ is just $\nc$. 

\begin{figure}[b!]
\begin{center}
\includegraphics[width=3.5in,trim=0.5in 0 0.4in 0,clip=true]{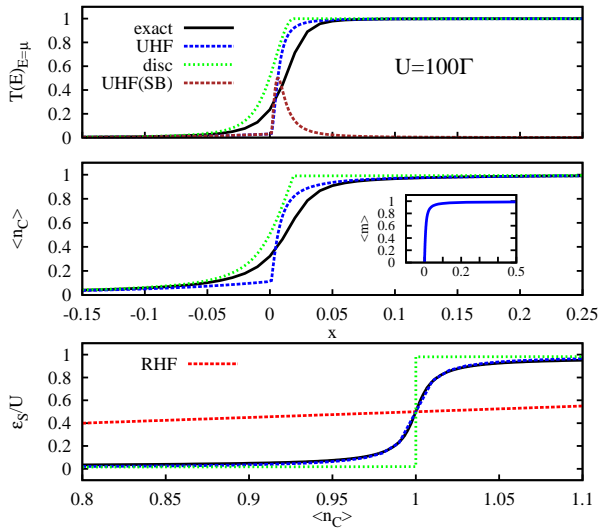}
\caption{Upper panel: transmission as a function of $x=(\mu-\epsilon)/U$; middle panel: occupation as a function of $x$; lower panel: KS potential as a function of occupation. Results are shown for Bethe ansatz or exact KS DFT (exact), unrestricted Hartree-Fock (UHF), and discontinuous approximation [disc, Eq. \eqref{ldau}]. Also shown in the upper panel is UHF results for transmission using (incorrect) spin densities [UHF(SB), with symmetry breaking], and $\left<m\right>=\left<n_\uparrow\right>-\left<n_\downarrow\right>$ for UHF as a function of $x$ as an inset in the middle panel. $U=100\Gamma$ in all cases, and only the region near $\nc=1$ and $x=0$ is shown.}
\label{f:u100g1}
\end{center}
\end{figure}

While this model does contain a discontinuity, and yields the exact result as $y\to 0$,
curing the worst defects of HF, it misses entirely the finite slope of the KS potential
at half-filling for finite $U$, which is determined by the susceptibility.  
The explicit derivative discontinuity is exact in the strongly-correlated limit with infinite
$U/\Gamma$, but should be ``rounded'' in finite $U/\Gamma$ \cite{BLBS11,ES11}, or in
finite temperature \cite{SK11}. To see this for finite (but very large) $U/\Gamma$, in  Fig. \ref{f:u100g1}, we show similar results as in Figs. \ref{f:u1g1} and \ref{f:u10g1}, but with $U=100\Gamma$ and we only show the region around $\nc=1$ at $x=0$, where the rounded derivative discontinuity occurs.
  The transmission is accurate both for weak and strong correlation,
but is not so everywhere in between.   In particular, it is overestimated for $\nc$ just
above 0 (and just below 1) for $U=10\Gamma$ because of this lack of a finite slope.
This is where we expect the greatest errors in such models, but the region of inaccuracy (on the scale of $x$) shrinks as $U/\Gamma \to \infty$.

Finally, we discuss a different class of approximations.
A well-known (and much debated) technique for mimicking strong correlation
is to allow a mean-field calculation to break symmetries that are preserved in the
exact calculation.  Perhaps the most celebrated prototype of such a calculation is
for HF applied to an H$_2$ molecule with a large bond distance.  At a crucial value
of the bond distance (called the Coulson-Fischer point), an {\em unrestricted} 
calculation, i.e., one that allows a difference in spin occupations, yields a lower
energy than the restricted one.  This remains the case for all larger separations,
and the unrestricted solution correctly yields the sum of atomic energies
as $R\to\infty$, whereas the restricted Hartree-Fock (RHF) solution dissociates to unpolarized H atoms with 
the wrong energies.   This is the celebrated symmetry dilemma:  with a mean-field
approximation, for large separations, one can either get the right symmetry (RHF)
or the right energy [unrestricted Hartree-Fock (UHF)], but not both.  The same issues arise in approximate
DFT
treatments of this problem \cite{PSB95}.  Of course, the exact functional
manages to get the correct energy with the correct symmetry, and there have been
many attempts to reproduce this with various more sophisticated approximations.
But a more pragmatic approach is to accept the results as they are, interpreting the
good energetics as the result of applying the approximate functional to a frozen
fluctuation of the system.  The true ground-state wavefunction fluctuates between
configurations with one spin and then the other (left and right localized for stretched
H$_2$), and the true ground-state density has unbroken symmetry.  But the approximate
functionals give most accurate energies when applied to the frozen fluctuations.
Thus, one can interpret both the {\em total} density and energy as being accurate
from such a calculation, but not the individual spin-densities.  In fact, 
an alternative approach is to interpret another variable, such as the ontop pair
density, as being accurately approximated in such treatments \cite{PSB95}.

We apply the same reasoning to the Anderson junction, just as was done by Anderson when
creating the model we are using \cite{A61}.  The symmetries are different, but the principle
is the same.  We allow the mean-field calculation to break spin-symmetry if this
leads to lower energy on the central site, with spin equations:
\ben
\left<n_\uparrow\right>=\frac{1}{2}+\frac{1}{\pi}\arctan \left(\frac{\mu-\varepsilon-U\left<n_\downarrow\right>-\varepsilon\xc\left(\left<n_\uparrow\right>,\left<n_\downarrow\right>\right)}{\Gamma}\right),
\een
and reverse for $\left<n_\downarrow\right>$, and $\nc=\left<n_\uparrow\right>+\left<n_\downarrow\right>$. 
Again, the simplest calculation is UHF, where $\varepsilon\xc=0$.  The solutions are identical
to those found in the original problem by Anderson \cite{A61}.   For $y > 1/ \pi$, i.e., $U < \pi
\Gamma$, there is no spontaneous symmetry-breaking, and UHF=RHF.  But beyond that
critical value, the spin-density difference becomes finite, and the unrestricted solution
differs.  Define the density difference as $\left<m\right>=\left<n_\uparrow\right>-\left<n_\downarrow\right>$ in UHF, which satisfies:
\ben
\tan\left(\frac{\pi}{2}\left<m\right>\right)=\frac{\left<m\right>}{2y}. \hspace{0.2in} \mbox{UHF}
\label{uhfndelta}
\een
For $y > 1/\pi$, $\left<m\right>=0$, but otherwise a solution with $\left<m\right>$ finite exists.
In all cases, we take only the total density from the UHF calculation, and
we know the true $\left<m\right>=0$ always.
In particular, as $y\to 0$ (strong correlation), $\tchic \to 0$
with the correct linear term:
\ben
\tchic \to (8/\pi) y+(96\gamma/\pi^2) y^2 +\cdots \hspace{0.2in} y \to 0,
\label{uhfexpand}
\een
where $\gamma=1$ in the exact solution, but $\gamma=1/3$ in UHF. So UHF recovers the leading term.  The green curve in Fig. \ref{fig:exactcond} shows 
the UHF value of $\tchic$, demonstrating both its accuracy for
both strong and weakly correlated systems, and the discontinuous change at $1/\pi$.

Even beyond the ``Coulson-Fisher point" of $1/\pi$, the symmetry-breaking
only occurs for $0 \leq \left<m\right> \leq 1$, i.e., outside this region, the UHF solution
is that of RHF, as can be seen in the inset of middle panel in Fig. \ref{f:u100g1}. But the density is very accurately given by UHF (considering the scale of horizontal axis), and
the KS potential develops the correct derivative discontinuity as $y\to 0$.

To demonstrate this accuracy, we plot the corresponding transmissions in Fig. \ref{f:u10g1}, using Eq. \eqref{tnospin}.
The figure shows how the transmission using $\nc$ from UHF is almost exact (considering the scale of horizontal axis).
To demonstrate the error in ignoring the fact that the UHF produces incorrect spin
densities, we also plot the transmission through such a solution, which is completely
wrong (see dark red curve in upper panel of Fig. \ref{f:u100g1}, only one peak is present because only region near $x=0$ and $\nc=1$ is shown there).  Our results are consistent with those of \cite{A61}, justifying the use of the
broken symmetry solution to deal with strong correlation.

To summarize, we have studied approximate treatments of the
zero-temperature weak-bias conductance of the
Anderson junction.
RHF and approximate DFT treatments work well for weak correlation,
but fail for moderate and strong correlation 
because of the smooth dependence of their KS potentials
on occupation numbers.
Imposing an explicit discontinuity consistent with particle-hole symmetry can yield a discontinuity with occupation
which guarantees correct behavior in the strong correlation limit.
This also greatly improves results for moderate correlation, but still contains errors.
Finally, simple symmetry-breaking in UHF produces remarkably accurate
conductances, once the transmission is calculated {\em as if} the symmetry had {\em not}
been broken.

KB acknowledges support from the Department of Energy under Award Number DE-FG02-08ER46496. CAS acknowledges support from the Department of Energy under Award Number {D}{E}{-}{S}{C}0006699. 

\bibliography{lit}
\end{document}